\documentclass[]{aastex631}

\newcommand{\narda}{2016\,SD$_{106}$}

\newcommand{\asiaa}{Institute of Astronomy and Astrophysics, Academia
  Sinica, No.1, Sec. 4, Roosevelt Rd, Taipei 10617, Taiwan}

\newcommand{\upenn}{Department of Physics and Astronomy, University of
  Pennsylvania, 209 S. 33rd St., Philadelphia, PA 19125, USA}

\usepackage{CJKutf8}
\usepackage{hyperref}
\usepackage[noabbrev,capitalise]{cleveref}

\makeatletter
\usepackage{etoolbox}
\patchcmd\H@refstepcounter{\protected@edef}{\protected@xdef}{}{}
\makeatother

\shorttitle{low-$i$ neutral extreme TNO}
\shortauthors{Chen et al.}

\graphicspath{{./}{figures/}}

\begin{document}

\title{A low-inclination neutral Trans-Neptunian Object in a extreme orbit}

\correspondingauthor{Ying-Tung Chen}
\email{ytchen@asiaa.sinica.edu.tw}

\author[0000-0001-7244-6069]{Ying-Tung Chen (\begin{CJK*}{UTF8}{bkai}
    陳英同\end{CJK*})}
\affiliation{\asiaa}

\author[0000-0002-0760-1584]{Marielle R. Eduardo}
\affiliation{\asiaa}

\author[0000-0002-0792-4332]{Marco A. {Mu{\~n}oz-Guti{\'e}rrez}}
\affiliation{\asiaa}

\author[0000-0001-6491-1901]{Shiang-Yu Wang (\begin{CJK*}{UTF8}{bkai}
    王祥宇\end{CJK*})}
\affiliation{\asiaa}

\author[0000-0003-4077-0985]{Matthew J. Lehner}
\affiliation{\asiaa}
\affiliation{\upenn}

\author[0000-0003-1656-4540]{Chan-Kao Chang (\begin{CJK*}{UTF8}{bkai}
    章展誥\end{CJK*})}
\affiliation{\asiaa}

\begin{abstract}
We present photometric observations and numerical simulations of \hbox{\narda},
a low inclination ($i=4.8\degr$) extreme trans-Neptunian Object
with a large semi-major axis ($a=350$\,au) and perihelion ($q=
42.6$\,au). This object possesses a peculiar neutral color of $g-r =
0.45\pm0.05$ and $g-i=0.72\pm0.06$, in comparison with other
distant trans-Neptunian objects, all of which have moderate-red to
ultra-red colors. A numerical integration based on orbital fitting on
astrometric data covering eight years of arc confirms that
\hbox{\narda} is a metastable object without significant
scattering evolution. Each of the clones survived at the end
of the 1\,Gyr simulation. However,
very few neutral objects with inclinations $<5\degr$ have been found
in the outer solar system, even in the main Kuiper belt.
Furthermore, most
mechanisms which lift perihelion distances are expected to produce a
very low number of extreme objects with inclinations $<5\degr$. We
thus explored the possibility that a hypothetical distant planet could
increase the production of such objects. Our simulations show that no 
\hbox{\narda}-like orbits can be produced from three Kuiper belt populations 
tested (i.e. plutinos, twotinos, and Haumea Family) without the presence of 
an hypothetical planet, while a few similar orbits can be obtained with it; 
however, the presence of the additional planet produces a wide range of 
large semimajor-axis / large perihelion objects, in apparent contradiction 
with the observed scarcity of objects in those regions of phase space.
Future studies may determine if there is a connection between the existence of a perihelion gap and a particular orbital configuration of an hypothetical distant planet.

\end{abstract}

\keywords{celestial mechanics --- surveys --- Kuiper belt objects:
 individual (2016 SD106)}

\section{Introduction} \label{sec:intro}
Trans-Neptunian Objects (TNOs) are not simply a group of planetesimals
that survived after the early evolution of the solar system, but are also a
complicated combination of remnants transferred from different
locations of the solar system during the planetary migration. TNOs can
be broadly classified into dynamically cold and hot populations. 
The dynamically cold objects (the so-called cold classical Kuiper Belt objects),
which observationally congregate at inclinations $i < 5\degr$ and around semi-major axes $a$ between 39.4 and
47.7\,au, are thought to be formed in situ and have survived after the
migration of Neptune, whereas many of the dynamically hot objects that
spread over a wide range of semi-major axes from 30 to many hundreds
au were perturbed/scattered by Neptune, i.e., hot classical Kuiper Belt objects, 
objects in mean-motion resonances (MMRs) with Neptune and scattered disc objects
(SDOs). Additionally, detached objects (DOs) are not in resonance with
Neptune, and typically have semimajor axes $a > 47.4$\,au,
eccentricities $e > 0.24$, and perihelion $q \ge 38$\,au. 
Because the exact boundary between SDOs and DOs are still being investigated, 
we refer to $q > 40$ and $a > 250$~au TNOs as ``extreme'' TNOs.
DOs are not influenced gravitationally by Neptune or the other currently known giant
planets (see the dynamical reviews by \citet{sai20} and
\citet{gla21}), and the physical mechanisms that placed the DOs into
their current orbits remain undetermined. Such objects likely
originated from the SDO population, but they are now totally
disconnected from it because their perihelia have been lifted in the
past by external perturbers or other unknown mechanisms. Several
scenarios to lift $q$ have been proposed: (1) gravitational influence
of nearby stars in the Sun's birth cluster \citep{bra06, bra12}, (2) a
passing star \citep{mor04, ken04}, (3) undetected planet(s) in the
past \citep{gla06, sil18}, or at present \citep{tru14, bat16, bat21}.
However, current data are insufficient to identify the dominant
mechanism.

Surface composition plays another key role in the investigation of the
origin of solar system objects. A correlation between color and
inclination was found for the population of classical TNOs
\citep{tru02}. The dynamically cold objects, i.e. cold classical
Kuiper Belt objects, which have redder colors, are representative of the
primordial bodies formed beyond Neptune's orbit. In contrast, the
dynamically hot objects, including hot classical objects, MMRs, SDOs,
and DOs, were formed in different regions of the solar system before
being excited during Neptune's migration, and thus have a wider color
distribution. For example, the color distributions of observed
plutinos and twotinos (3:2 and 2:1 MMR with Neptune) range from
neutral to ultra red. We note that the MMRs and SDOs often experience
resonance sticking or Kozai interactions, which change the semi-major
axes, eccentricities, and inclinations of TNOs and/or switch them to
other resonances, but both interactions are rarely seen for objects
with $a > 250$\,au. In addition, the Haumea family, the well-known
collisional population, has a unique neutral color distribution ($g-r
< 0.5$), which is different from the color of other populations.
Their neutral surfaces possibly originate from their past collisional
history which exposed underlying material \citep{tru07, rab08}.
Subsequent discoveries with similar surface colors and inclinations
support this hypothesis \citep{brown07}. To date, very few neutral
objects in the outer solar system have been confirmed with $i < 5\arcdeg$.

The Hyper Suprime-Cam Subaru Strategic Program
(HSC-SSP)\footnote{\url{https://hsc.mtk.nao.ac.jp/ssp/}} is a deep
multi-band imaging survey of 1400\,deg.$^2$ of the sky with the 8-m
Subaru telescope \citep{aihara18}. Although the main science goals are
cosmology and galaxy evolution, its depth and field coverage provide a
unprecedented dataset to detect distant/faint objects beyond
Neptune. With this dataset, we have discovered 178 TNOs in Deep/UDeep
fields (Eduardo et al., in preparation) and 231 TNOs in Wide fields \citep{chen18} including
an extremely neutral DO. Because the cadence of HSC-SSP was not
designed for color observations of TNOs, corrections for rotational
and phase angle effects could not be made. We thus made additional
observations to obtain precise multi-color measurements.

In this study, we present the color estimation of this extreme neutral
object, along with the results of simulations of \hbox{\narda}, (with
and without a hypothetical additional planet) to explore its stability
and possible origin.

\section{Observations} \label{sec:obs}

The Dark Energy Survey (DES) team discovered \hbox{\narda} and first
submitted the observations to the Minor Planet Center (MPC)
\citep{bern22}. However, \hbox{\narda} was also independently
discovered in the HSC-SSP DEEP2-3 Deep field. Data from the HSC-SSP
Deep fields include repeated multi-band ($grizy$) observations at
different epochs. The number of measurements in $g$, $r$, and $i$ are
10, 10, and 14, respectively. The HSC-SSP typically finished all
observations for a single filter for Deep/UDeep fields within one
night, then observations for other filters for the same field usually
were executed in the same dark run or later runs. The color estimation
of \hbox{\narda} from HSC-SSP observations shows a neutral color with
a very large uncertainty ($g - r = 0.50 \pm 0.24$ and $g - i = 0.70
\pm 0.23$). However, most moderate/small size TNOs are known to be
non-spherical with rotation periods in the range of several hours. The
observed cross-sections thus change with time, causing inaccurate
color estimates if observations using different filters are not nearly
simultaneous. Therefore, we requested and received CFHT Director's
Discretionary Time (DDT) to confirm the color estimate of
\hbox{\narda}. Consecutive observations for color measurements were
completed using MegaCam \citep{bou03} on CFHT within a span of 1.8 hours on
2021 November 2. The observations were conducted using a sequence of
filters of the form
$r$\text{-}$g$\text{-}$r$\text{-}$i$\text{-}$i$\text{-}$r$\text{-}$z$\text{-}$r$\text{-}$g$\text{-}$r$\text{-}$i$\text{-}$i$\text{-}$r$\text{-}$z$\text{-}$r$, with all observations performed using
sidereal tracking. Assuming a uniform surface color and linear
variation, this sequence would remove any rotation effects. The
astrometry and photometry of the CFHT DDT and HSC-SSP data were
calibrated using the Pan-STARRS1 catalog \citep{sch12, tonry12,
  mag13}. The photometry was re-measured with the moving-object
photometry package TRIPPy \citep{fra16} using a point spread function
of reference stars to generate trailing apertures according to object motion.

\begin{figure*}[ht!]
\centering
\plotone{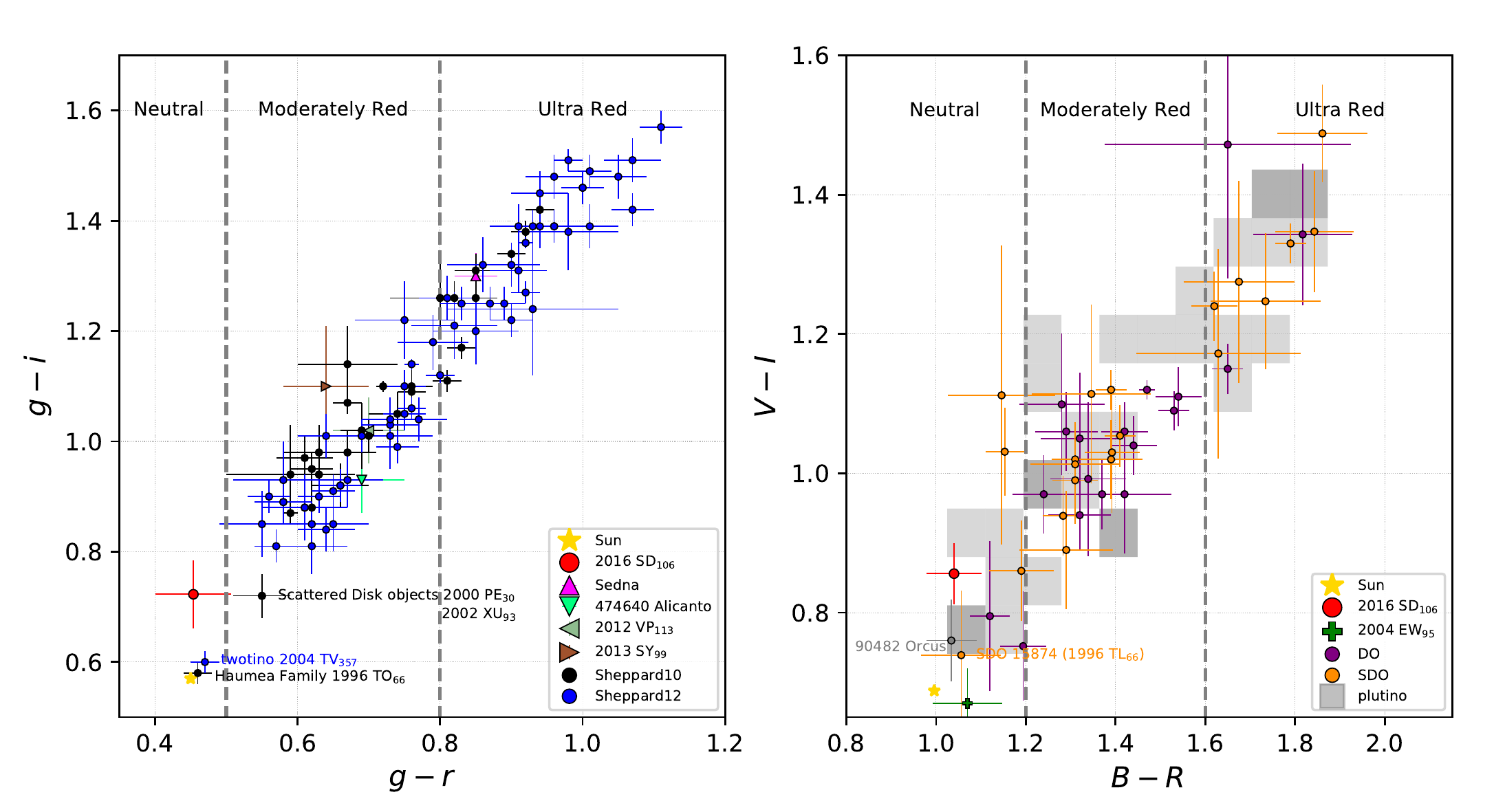}
\caption{Left panel: $gri$ colors of \hbox{\narda}, selected extreme TNOs, and objects from \citet{she10} and
  \citet{she12}, which include most of the dynamically excited TNO
  populations. The colors of high-perihelion objects mentioned in
  \citet{ban17} are indicated by triangles, and the solar color is
  indicated by the yellow star. The dashed lines roughly indicate the
  different color regions. Right panel: $BVRI$ colors of \hbox{\narda},
  the particularly neutral plutino 2004\,EW95 \citep{per13}, and
  plutinos, SDOs, and DOs from MBOSS \citep{hai12}.
  The $BVRI$ color of \hbox{\narda} was transformed
  from the original Sloan colors using the equations from
  \citet{smith02}.}
\label{fig:nardacolor}
\end{figure*}

Upon completion of the analysis, we measured the color of
\hbox{\narda} to be $g - r = 0.45 \pm 0.05$ and $g - i = 0.72 \pm
0.06$, indicating an extreme neutral surface (see
\Cref{tab:tab_color}) similar to those of neutral resonant objects
(2004\,EW$_{95}$, 90482\,Orcus, and 2004\,TV$_{357}$), SDOs
(1996\,TL$_{66}$), the solar $g - r = 0.45 \pm 0.02$ \citep{hol06},
and the Haumea Family member (1996\,TO$_{66}$) $g-r \sim 0.46\pm0.02$
(see \Cref{fig:nardacolor}). Comparing with the color database
in Minor Bodies in the Outer Solar System \citep[MBOSS,][]{hai12},
\hbox{\narda} is the bluest object among all known SDOs/DOs with color measurements. If the peculiar color of
\hbox{\narda} implies a high albedo of $p \simeq 0.51$, as the Haumea
Family \citep{ort17}, the estimation of size is $\sim80$\,km; the
estimation of size is $\sim250$\,km with general low albedo of $p =
0.05$. Although we do not know the variability of the phase-curve and
light curve, the absolute magnitude could be roughly estimated as
$H_{r} = 6.46$. We note that the $g - r$ (0.45) agrees to the other
known neutral objects, but $g - i$ (0.71) is higher than typical value
($\sim0.6$) due to the brighter magnitude on $i$-band. Following the
same method in \citet{she10}, the spectral gradient (SG) based on $g -
r$ and $g - i$ is 1.1 and 6.1 percent of reddening per 100\,nm,
respectively. (We are not reporting the $z$-band photometry because
there are currently too few DOs with such measurements for useful comparisons.)

To minimize the orbital uncertainties, we combine the observations
from HSC-SSP, CFHT DDT and
DES\footnote{\url{https://minorplanetcenter.net/db_search/show_object?object_id=2016+SD106}}
for orbital determination. Based on the observations spanning over 8
oppositions, we determined the orbit using the orbit fitting code of
\citet{ber00}. The resulting uncertainties in heliocentric orbital
elements are all small: $a = 354.23\pm0.49$\,au, $e =
0.87973\pm0.00014$, $i = 4.8080\pm0.0001 \degr$, longitude of
ascending node $\Omega = 219.494\pm0.002 \degr$, and argument of
pericenter $\omega = 163.032\pm0.005$. Pericenter passage will occur
at $2464675.363\pm0.84$.

\begin{deluxetable*}{cccccccclcl}
\centering
\tablecaption{Orbital elements and colors of selected extreme TNOs and
 representative objects with neutral color in the outer solar
 system.\label{tab:tab_color}}
\tablecolumns{10}
\tablehead{
\colhead{Designations} &
\colhead{$q$} &
\colhead{$a$} &
\colhead{$e$} &
\colhead{$i$} &
\colhead{$H$} &
\colhead{Arc} &
\colhead{$g-r$} &
\colhead{$g-i$} &
\colhead{Orbit} &
\\
\colhead{} &
\colhead{(au)} &
\colhead{(au)} &
\colhead{} &
\colhead{($\degr$)} &
\colhead{} &
\colhead{(days)} &
\colhead{} &
\colhead{} &
\colhead{Class} &
}
\startdata
\hbox{\narda} & 42.65 &354.7 $\pm$ 0.4 & 0.87 & 4.80 & 6.8 & 2977 &
0.45 $\pm$ 0.05 & 0.72 $\pm$ 0.06 & extreme\\
90377\,Sedna & 76.36 & 510.3 $\pm$ 0.1 & 0.85 & 11.93 & 1.55 & 11418 &
0.85 $\pm$ 0.03 & 1.31 $\pm$ 0.04\tablenotemark{a} & extreme\\
474640\,Alicanto & 47.29 & 344.4 $\pm$ 1.1 & 0.86 & 25.53 & 6.46 &
7736 & 0.69 $\pm$ 0.06 & 1.01 $\pm$ 0.06\tablenotemark{a} & extreme\\
2012\,VP$_{113}$ & 80.43 & 269.2 $\pm$ 0.5 & 0.70 & 21.07 & 4.09 &
3028 & 0.70 $\pm$ 0.05 & 1.02 $\pm$ 0.06\tablenotemark{b} & extreme\\
2013\,SY$_{99}$ & 50.08 & 824 $\pm$ 32 & 0.93 & 4.21 & 6.7 & 1156 &
0.64 $\pm$ 0.06 & 1.10 $\pm$ 0.11\tablenotemark{c} & extreme\\
\cline{1-10}
19308\,(1996\,TO$_{66}$) & 38.49 & 43.528 $\pm$ 0.001 & 0.11 & 27.33 &
4.88 & 13933 & 0.46 $\pm$ 0.02 & 0.58 $\pm$ 0.02\tablenotemark{a} &
Haumea\\
120216\,(2004\,EW$_{95}$) & 26.97 & 39.429 $\pm$0.008 & 0.35 & 29.31 &
6.61 & 7035 & 0.47 $\pm$ 0.06 & 0.53 $\pm$ 0.09\tablenotemark{d*} &
3:2\\
90482\,Orcus & 30.13 & 39.0971 $\pm$0.0004 & 0.23 & 20.57 &
2.19 & 25690 & 0.46 $\pm$ 0.04 & 0.64 $\pm$ 0.06\tablenotemark{f*} &
3:2\\
2004\,TV$_{357}$ & 34.50 & 47.715 $\pm$ 0.002 & 0.27 & 9.76 & 6.9 &
5607 & 0.47 $\pm$ 0.02 & 0.60 $\pm$ 0.02\tablenotemark{e} & 2:1\\
15874\,(1996\,TL$_{66}$) & 34.97 & 83.733 $\pm$ 0.003 & 0.58 & 23.95 &
5.46 & 9222 & 0.48 $\pm$ 0.07 & 0.67 $\pm$ 0.10\tablenotemark{f*} &
SDO\\
\enddata
\tablenotetext{}{\textbf{Note:} Except \hbox{\narda}, all orbital
  elements were taken from:
  \url{https://ssd.jpl.nasa.gov/tools/sbdb_lookup.html}.  For DOs, we
  only selected $q>40$ objects which have $gri$ color measurements in
  the literature. Adapting the transformation equations between $BVRI$ and
  $gri$ in \citet{smith02}, the $BVRI$ colors of \hbox{\narda} are
  given by $B-V = 0.66$ , $V-R = 0.37$, and $R-I = 0.47$. The * symbol
  indicates the measurements were transferred from $BVRI$ data.}
\tablenotetext{}{\textbf{References:} $a$ = \citet{she10}, $b$ =
  \citet{tru14}, $c$ = \citet{ban17}, $d$ = \citet{per13}, $e$ = 
  \citet{she12}, $f$ = \citet{hai12} }
\end{deluxetable*}

\section{Numerical Integration and Analysis}\label{sec:analysis}
\begin{figure}[ht!]
\centering
\plotone{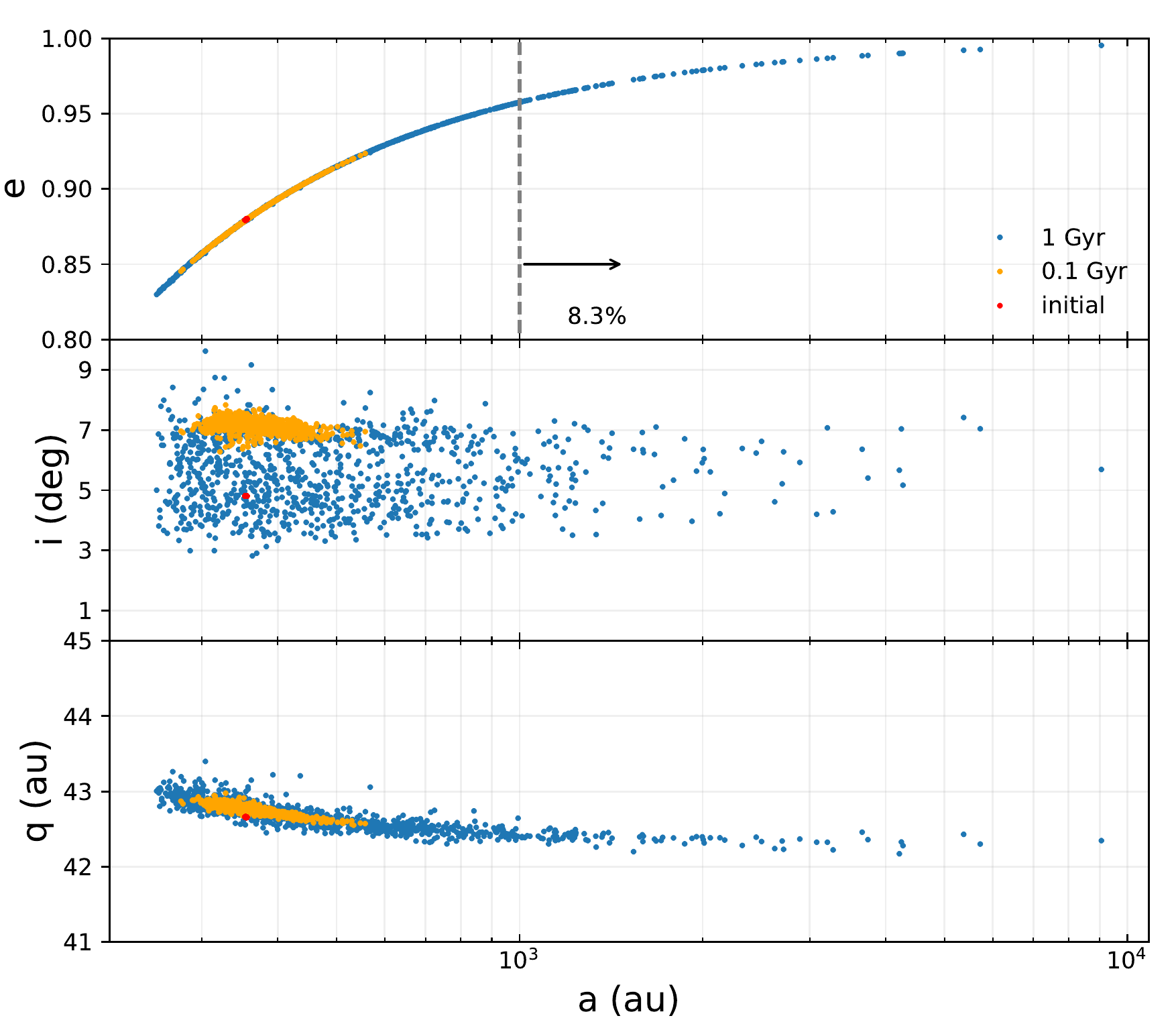}
\caption{The orbital distribution of 1,000 clones of \hbox{\narda}
  after 0.1\,Gyr (orange dots) and 1\,Gyr simulation (blue dots). The
  red dots indicate the initial conditions. The clone orbits are
  much more stable than typical Centaurs/SDOs, with $\Delta q <
  1$\,au over time. Only a few percent of clones evolve to $a >
  1,000$\,au. }
\label{fig:clones}
\end{figure}

With the available astrometric data over eight oppositions, we can
accurately estimate the orbital parameters of \hbox{\narda} and
determine its long-term stability. Using the covariance matrix output
from the orbit fitting code of \citet{ber00}, we generated
1,000~clones within 3$\sigma$ of \hbox{\narda} best-fit
orbit. We then performed 1\,Gyr long numerical
simulations using the symplectic integrator from the \texttt{MERCURY}
package \citep{cha99}, considering the four giant planets, a 180\,day
initial time-step, and a 10,000\,au ejection distance. We found that
all clones survived after 1\,Gyr, with 91.7\% of them keeping $a <
1,000$\,au. The orbital distributions of the clones at three different
times (0, 0.1, and 1.0 Gyr) are illustrated in
\Cref{fig:clones}. Our results indicate that a diffusive behavior
in $a$ dominates the evolution of \hbox{\narda}, which is similar to
the behavior of 2013\,SY$_{99}$ \citep{ban17}. The semi-major axes
diffused both inward and outward due to planetary perturbations. The
variations in inclination were small on a Gyr timescale, with an
oscillating period of $\sim$100\,Myr between $4\degr$ and $7\degr$.
The change of perihelion distances was very small over time ($\Delta q <
1$\,au), compared, e.g. with the few au change found for the
semi-stable extreme object 2000\,CR$_{105}$ \citep{gla02}.

\begin{figure*}
  \plottwo{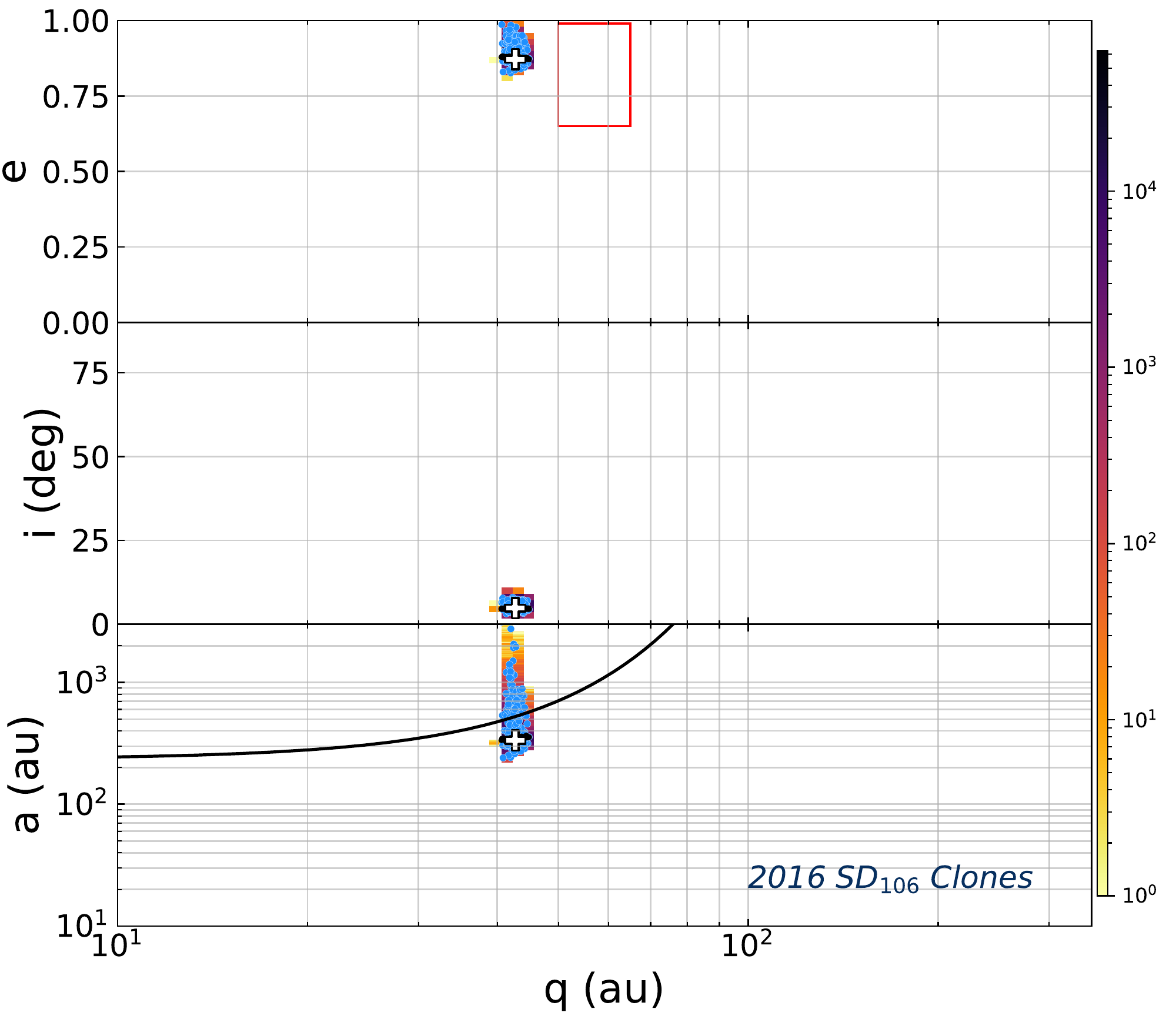}{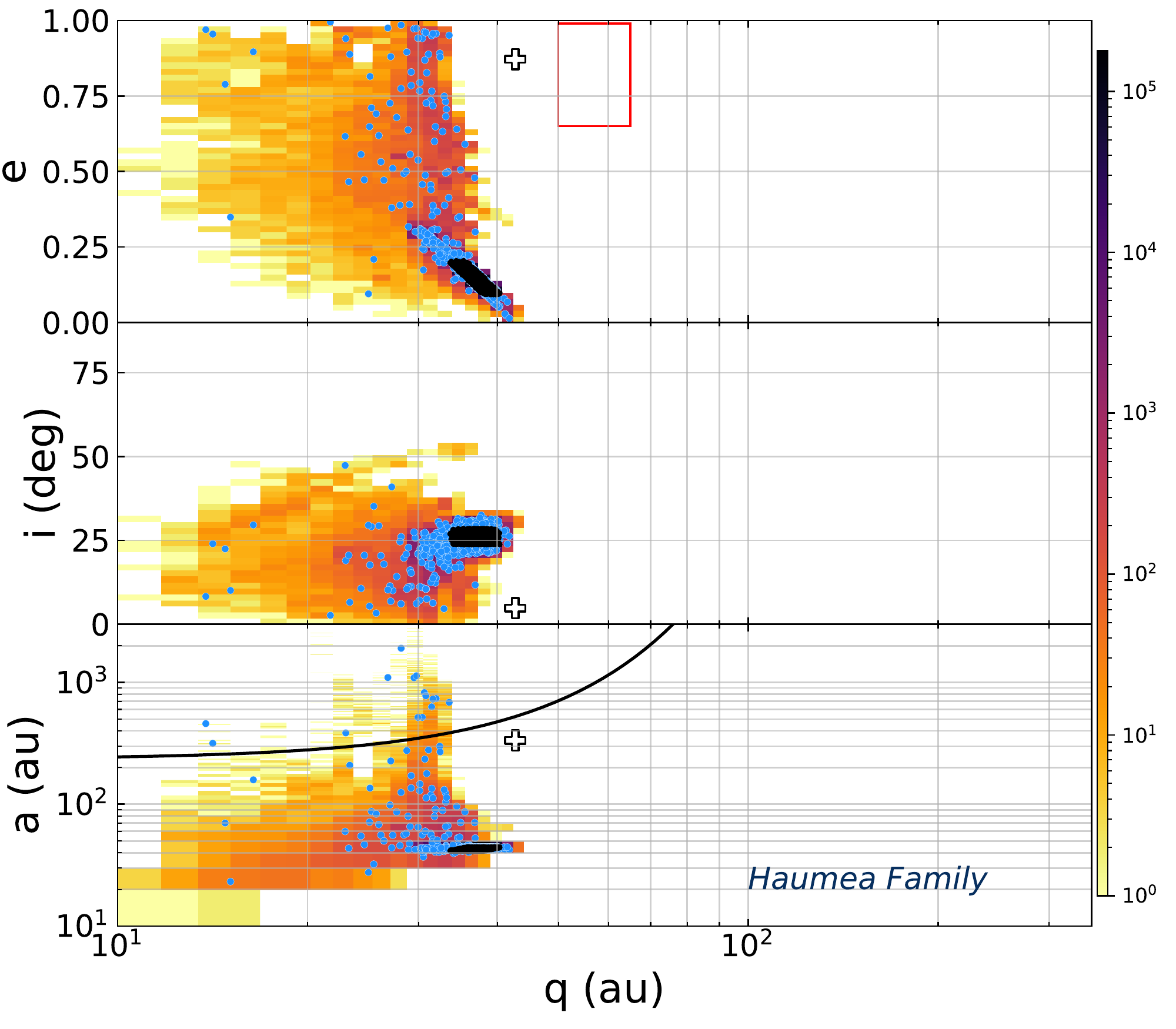}
  \plottwo{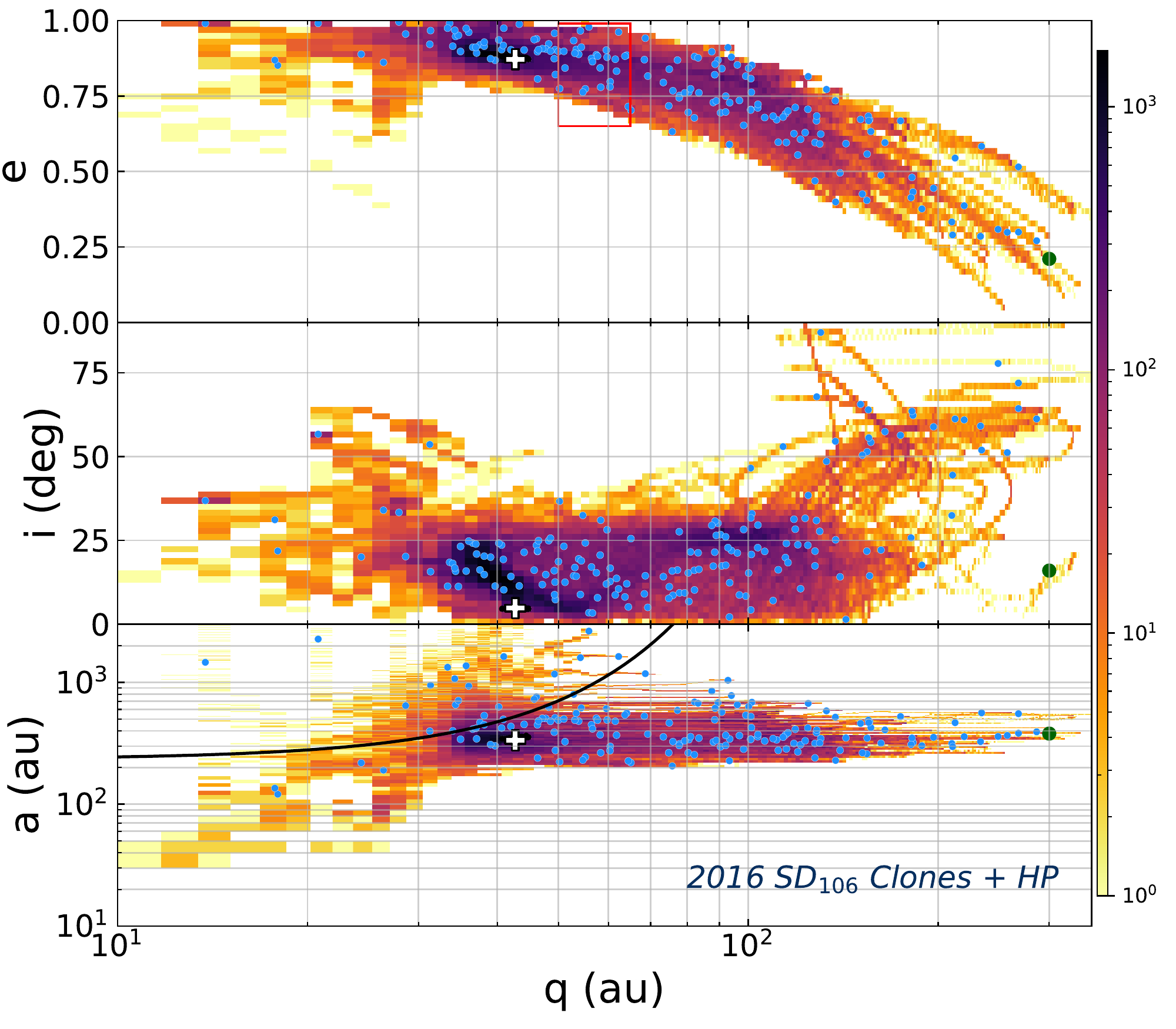}{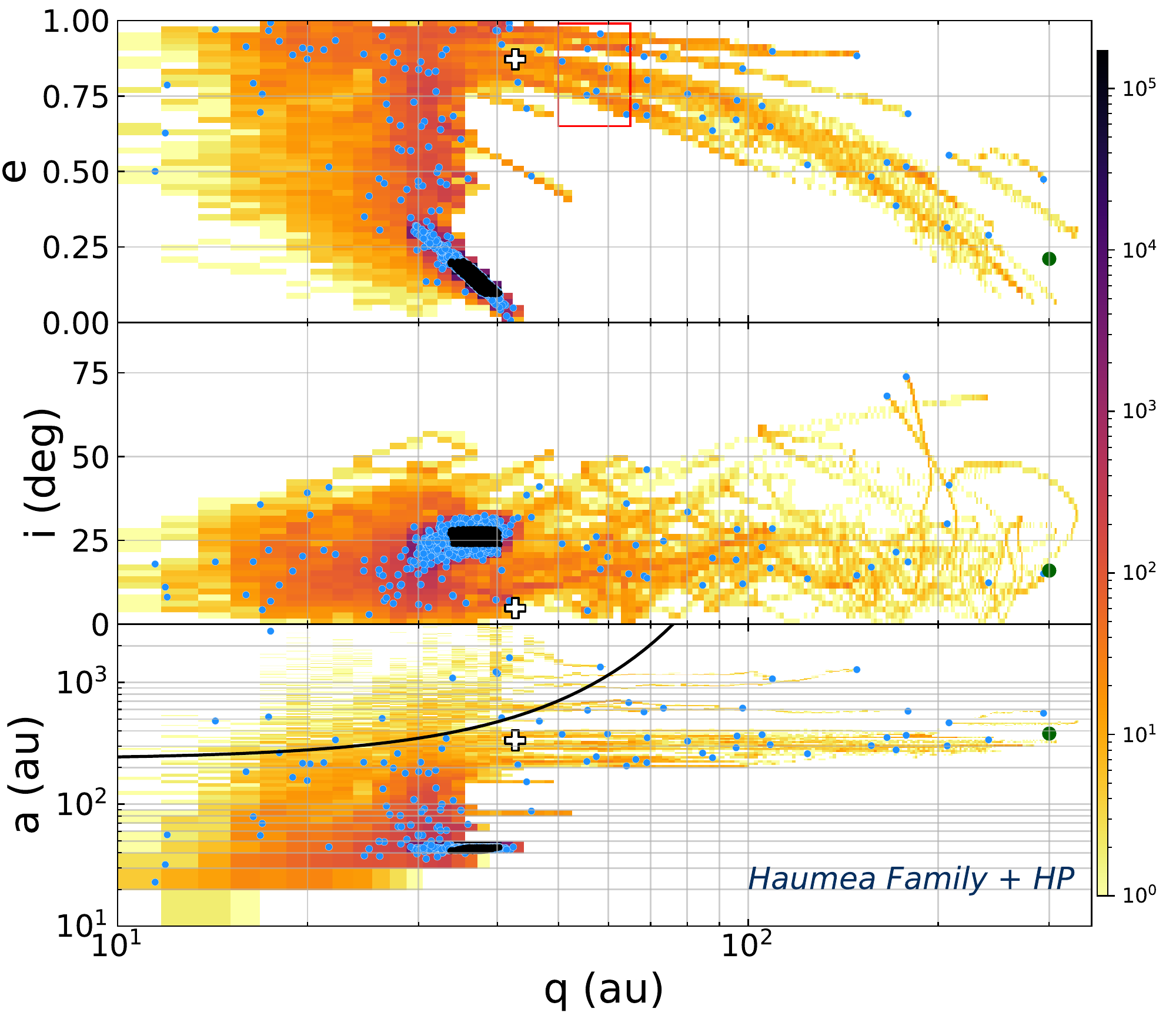}
\caption{Heatmaps of the orbital evolution of \hbox{\narda} clones and
  Haumea Family members in simulations with and without the HP.  The
  white cross and green dot show the initial orbit of \hbox{\narda}
  and HP, respectively. The initial orbits in each simulation are
  overplotted as black dots, while light blue dots show the final
  distribution of surviving particles after 1~Gyr. The observational
  $q$-gap suggested by \citet{old21} is outlined with a red
  square. The black curve shows the stability limit for distant SDOs as
  derived by \citet{bat21}; above this line in the $a$-$q$ plane,
  particles are quickly destabilized and ejected from the solar
  system.
\label{fig:narda_n_haumea}}
\end{figure*}

\begin{figure*}
\plottwo{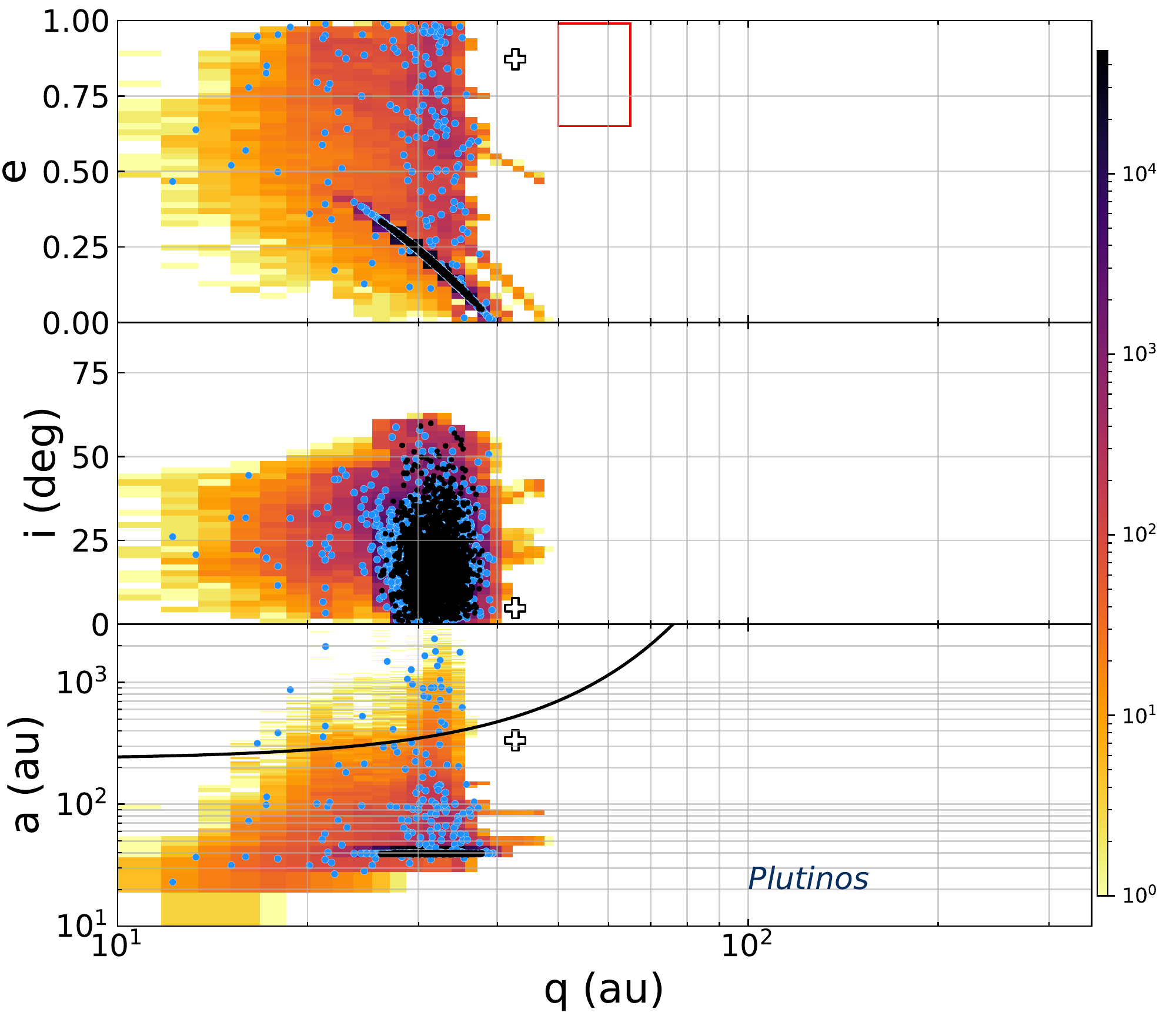}{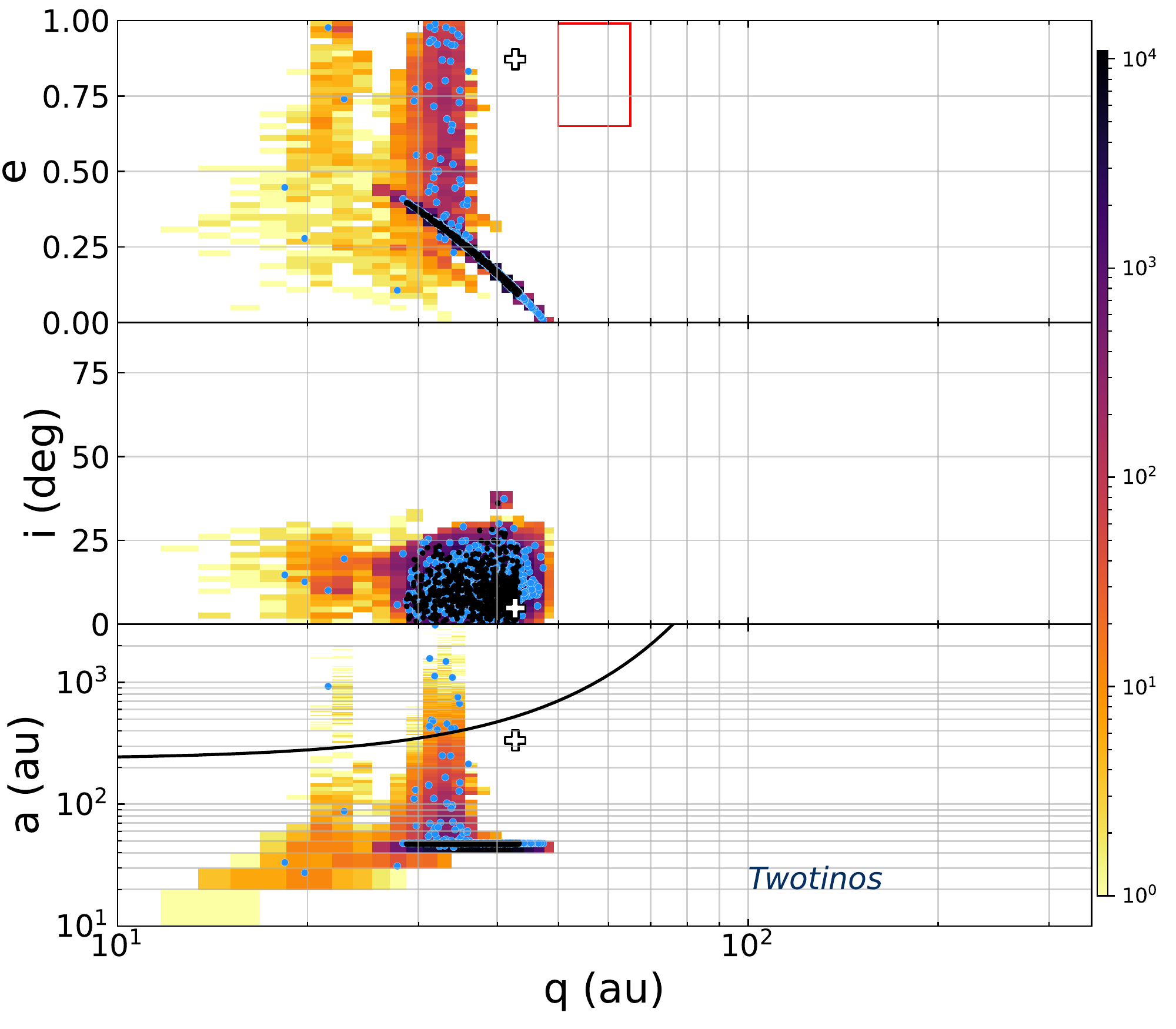}
\plottwo{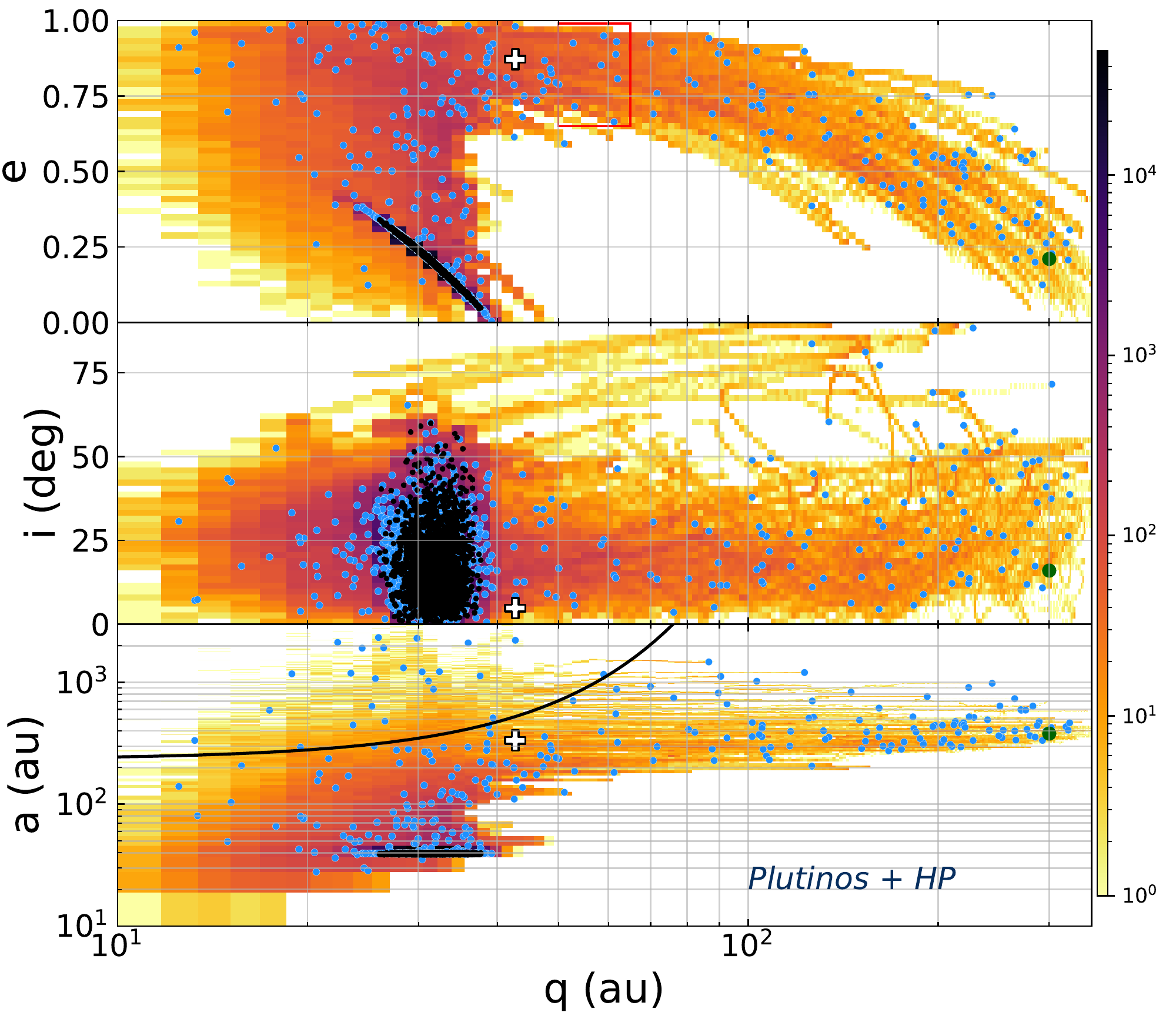}{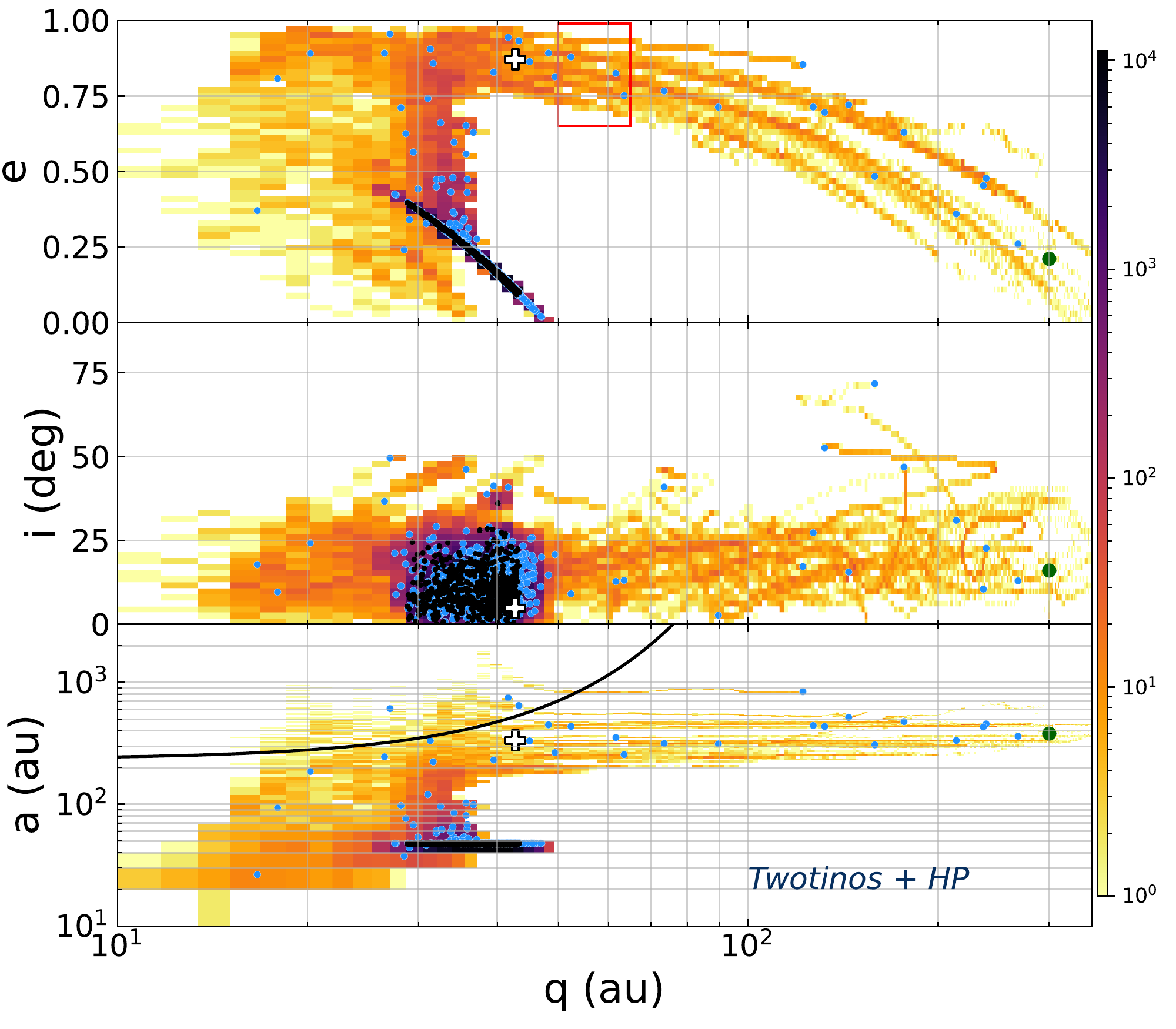}
\caption{Same as \Cref{fig:narda_n_haumea} except for plutinos
  and twotinos.
\label{fig:32_n_21}}
\end{figure*}

The semi-major axis of \hbox{\narda} is quite close to that of the
hypothetical planet (HP) of \citet{brown21}, and the existence of this
HP could dramatically change the orbital elements of distant
TNOs \citep{sha17, ban17}. In order to explore possible origins of
low-$i$ neutral DOs, we implemented experimental
simulations for four dynamical populations with/without the HP from
\citet{brown21}: (1) \hbox{\narda} clones in this study, (2) plutinos,
(3) twotinos, and (4) Haumea family members, as these populations
contain known neutral objects. Plutino and twotino particle
distributions were taken from the L7 model of \citet{pet11}. We used
the objects identified as resonant (those showing libration in a
10\,Myr interval) in \citet{mun19}, totalling 3340~plutinos and
870~twotinos. For the Haumea family, we generated a random set of
2000~objects, within the boundaries 42\,au $< a < 44.5$\,au, $0.1 < e
< 0.2$, and $24\degr < i < 28.5\degr$, following \citet{lyk12}. The
arguments of perihelion, longitudes of ascending node, and mean
anomalies, were randomly chosen between $0\degr$ and $360\degr$ in all
cases. We used \texttt{MERCURY}, with an increased initial timestep of 400\,d (in order to speed up integrations without loss of precision) and an
ejection distance of 20,000\,au, to integrate our systems for
1\,Gyr. We considered the evolution of particles under the influence
of the giant planets, with and without the presence of the HP ($a = 380$ au, $e=0.21$, $i=16\degr$, $\omega=150^\circ$, $\Omega=100^\circ$, M$\;= 6.2$ M$_{\earth}$)
(\Cref{fig:narda_n_haumea,fig:32_n_21}).
We note that the longitude of perihelion of \hbox{\narda} ($\varpi \sim 22\degr$) lies within the predicted cluster of $\Delta\varpi$, given the orbital configuration of the HP used here.

With and without the HP, over 85\% of the surviving plutinos and twotinos remain within the range of $\Delta a_0 \pm 0.7$\,au after 1\,Gyr, and a similar fraction of the surviving Haumea family members remain in the range $42 < a < 44.5$\,au.
The \hbox{\narda} clones with no HP show very little change, but
the HP clearly has a significant influence on the evolution of the
\hbox{\narda} clone orbits. The $i$ and $q$ of particles from the
other dynamical populations evolved and spread evenly between $3\degr$
to $40\degr$ and 30 to 380\,au by the end of the simulation. 
We note that a population with large $i$ and large $q$ would be significantly harder to detect since the objects spend most of their time at high latitude and larger
barycentric distances. Therefore a large population of such objects would be required to result in the detection of \hbox{\narda}.
Our results show that the HP
does make it possible for particles in various TNO populations to
diffuse towards \hbox{\narda}-like orbits. However, such diffusion produces a significant population of large $a$ / large $q$ objects, in apparent contradiction with an identified observational scarcity of objects in this region of the solar system phase-space, i.e. the so-called perihelion gap \citep{old21}.

\section{Discussion} \label{sec:discussion}
The existence of low-$i$ neutral extreme objects has significant implications.
The DOs have $a$ and $i$ distributions similar to those of SDOs,
except the values of $q$ are large enough to avoid the strong
perturbations from Neptune. Scattering processes cannot vary $i$ much
since $q$ and the Tisserand parameter relative to Neptune, $T_\mathrm{N}$,
are almost constant during the scattering process
\citep{sai20}. Both SDOs and DOs have similar color distributions as
shown in \citet{hai12} and \Cref{fig:nardacolor}. The orbital boundary
between SDOs and DOs is not clearly defined, as it may vary by
different classification criteria, e.g. \citet{kha20}, \citet{gla21}
or a stability validation. For example, the well-known neutral SDO,
136199\,Eris was classified as DO in MBOSS. However, even if we neglect the
classification definition, TNOs still seem to follow the color-inclination 
correlation. The color-inclination
correlation demonstrated in \citet{pei15} shows a lack of objects
with ($i < 5\degr$, $B - R \sim 1.0$), and remains a weaker correlation for all TNOs.
The only known object in this study
close to these conditions is the plutino 612029 (1995\,HM$_{5}$), with
$i = 4.8\degr$ and $B - R = 1.01 \pm 0.2$. \citet{mar19} combined the
OSSOS dataset with previously published measurements (including
samples in \citet{pei15}), and investigated the spectral slope (SS) of
dynamical hot populations ($i > 5\degr$). We note that the definition
of the spectral slope (SS) here is different from the spectral
gradient (SG) we evaluated for \hbox{\narda}, although the two values are
comparable. In the detached group, only 2003\,FZ$_{129}$ ($a =
61.4$\,au, $e = 0.38$, $i = 5.8\degr$, $q= 38.0$\,au) shows a
moderate-red-to-neutral slope (9.88) and an inclination close to
$5\degr$, although it shows a moderate-red color $B- R = 1.32 \pm
0.04$. The inclinations of other DOs are larger than
$17\degr$. The few neutral objects with SS $< 5$ in
both SDOs/DOs all show medium to high inclinations (1996\,TL$_{66}$:
$23\degr$, 2000\,YC$_{2}$: $18\degr$, 2000\,PE$_{30}$:
$18\degr$). \hbox{\narda} is indeed an unique object among the known
SDOs and DOs. Although the inclination distribution
of the distant TNOs is poorly known due to relatively small numbers
and survey biases, most known SDOs/DOs do not cluster on the ecliptic
plane, considering that most outer solar system surveys have a bias
towards detecting low-$i$ objects. So the existence of \hbox{\narda}
constrains the mechanisms by which extreme objects are produced, and may imply that a particular dynamical evolution
occurred with a few extreme objects, producing the special surfaces/orbits which
don't follow the color-inclination correlation found in TNOs.

The following are possible scenarios related to the origin of \hbox{\narda}:
\begin{description}
\item[Kozai resonance in MMR] Considering the Kozai resonance as a mechanism to
  raise perihelia, Kozai interactions must be coupled with
  increases of both ${q}$ and ${i}$. It is difficult to explain low-i
  DOs with large $q$. Results in \Cref{sec:analysis} show that the
  possibility of \hbox{\narda} coming from a high order MMRs with
  Neptune is extremely low. The closest possible MMRs to
  \hbox{\narda} are 1:40 ($\sim351$\,au) and 1:41 ($\sim357$\,au),
  which are too weak to preserve long-lived resonant objects.
  Some
  literature also indicate that the resonant mechanism is too weak
  to produce DOs beyond 250\,au, especially for low-$i$ objects
  \citep{gal12, bra15}. Thus the Kozai resonance cannot explain the
  low $i$ and high $q$ of \narda. Recent studies also indicate
  that theoretical models of planetary migration are unlikely to
  produce high-perihelion ($q>40$), low-$i$ DOs
  \citep{kai16, pike17b}. In addition, the major MMRs beyond Neptune,
  i.e. plutinos and twotinos, are very stable at present, and Kozai resonance on these major MMRs is
  unlikely to produce DOs with only the four known giant planets (see
  \Cref{fig:32_n_21}).
\item[Object diffused from the inner fringe of the Oort cloud] The
  orbital elements indicate that \hbox{\narda} is in the diffusive
  region and away from the strong scattering region \citep{dun87,
  ban17}. 
  2013\,SY$_{99}$, with a larger
  semi-major axis $a = 824$\,au, could be explained with a diffuse
  behavior of the ``inner fringe'' of the Oort cloud in a long-time integration
  \citep{ban17} for $45<q<50$\,au and $1000<a<2500$\,au. The diffusion mechanism along with
  planetary perturbations could produce the population of
  2013\,SY$_{99}$, initiating from part of SDOs without an additional
  source of mass. The efficiency of this scenario is
  still unknown, which is critical to understand if the production
  of extreme neutral objects from SDOs is possible.
\item[Analogues to the Haumea family] The Haumea family is a
  collisional population which share similar orbital parameters and
  surface colors. The possible orbital spaces of this family, assuming a reasonable range of impact velocities, don't cover the region of the
  detached population \citep{lyk12}, and most of survived particles
  have $i = 20\sim35\degr$ and $a = 35\sim55$\,au. The currently
  known/possible Haumea family members have moderate inclinations ($i
  \sim 28\degr$), rather than low inclinations. If an extremely rare
  collision event happened in the DO region, then neutral
  objects could be formed in situ, in which case we should expect more
  neutral DOs to be detected.
\item[Hypothetical Planet] Our simulations with a HP showed the
  possibility to provide a path to bring TNOs
  to the detached region. This pathway also fills up the observational
  $q$-gap ($50 \lesssim q \lesssim 65$ au and $0.65 \lesssim e < 1$)
  suggested by \citet{old21}, who also point out that
  the presence of a HP would produce observable objects within
  the $q$-gap. 
  Different q-gap outcomes are produced with different orbital configurations of the HP, which suggests that a relationship between various dynamical populations, the observational q-gap depth, and an HP may exist.
  Some plutinos, twotinos, and Haumea family members can
  be influenced by the HP and approach the detached region and inner
  Oort Cloud. The $i$ and $q$ evolution of these clones lowers their
  detectability dramatically, like 2013\,SY$_{99}$ in \citet{ban17},
  as the clones spend almost all of their time at higher inclinations
  and larger perihelia.
\item[2004\,EW$_{95}$] Although the high-$i$ plutino hints at a
  different dynamical evolution, 2004\,EW$_{95}$ ($i = 29\degr$) is
  the first TNO confirmed with a neutral surface possibly composed by
  more carbon and silicates (ferric oxides and phyllosilicates) than
  typical TNOs \citep{sec18}. 2004\,EW$_{95}$ is thought to have
  originated closer to the Sun during planetary migration. This object
  indeed supports the model prediction that a small fraction of
  objects with carbonaceous surfaces could be resident
  in the outer solar system. Future spectral observations of
  \hbox{\narda} in optical and IR are essential to confirm its surface
  composition and albedo.
\end{description}

A search of the MPC Database for objects with $a > 250$ and $q > 40$
found that 2013\,RA$_{109}$ and 2014\,WB$_{556}$ are potentially neutral
color objects as well, although their inclinations are moderately high
($12.4\degr$ and $24.2\degr$).
We emphasize that there may be no significant discrepancy between the neutral color of \hbox{\narda} and the color-inclination correlation, since this correlation has not yet been fully measured over the different dynamical populations. While the color and orbital
elements of \hbox{\narda} provide valuable tracers of DO formation, more systematic color observations
of DOs discovered in future surveys (e.g. LSST) could
provide more clues to understand their formation.

\section{Summary} \label{sec:cite}
Our observations show the low-$i$ extreme TNO \hbox{\narda} has an
unusual neutral color of $g - r = 0.45 \pm 0.05$ and $g - i = 0.72 \pm
0.06$. A numerical integration of clones shows that its orbital
evolution over 1\,Gyr is stable, with some diffusion behavior. Current
dynamical populations with neutral surface members could not easily
produce \hbox{\narda}-like objects when only accounting for perturbations due to
known planets. Although a HP could provide the possibility for plutinos,
twotinos or Haumea family members to migrate towards the extreme
region, this mechanism would require a vast population to produce the
detection of \hbox{\narda}. This rare neutral object with large-$a$,
large-$q$, and low-$i$ provides an additional tracer to understand the
origin of extreme objects.

\begin{acknowledgments}
Based on data collected at the Subaru Telescope and retrieved from the HSC data archive system, which is operated by the Subaru Telescope and Astronomy Data Center (ADC) at NAOJ. 
Also based on observations obtained with MegaPrime/MegaCam, a joint project of CFHT and CEA/DAPNIA, at the Canada-France-Hawaii Telescope (CFHT) which is operated by the National Research Council (NRC) of Canada, the Institut National des Science de l'Univers of the Centre National de la Recherche Scientifique (CNRS) of France, and the University of Hawaii. The observations at the Canada-France-Hawaii Telescope were performed with care and respect from the summit of Maunakea which is a significant cultural and historic site.  
This publication uses data generated via the Zooniverse.org platform, development of is which funded by generous support, including a Global Impact Award from Google, and by a grant from the Alfred P. Sloan Foundation.
The Hyper Suprime-Cam (HSC) collaboration includes the astronomical communities of Japan and Taiwan, and Princeton University.
The HSC instrumentation and software were developed by the National Astronomical Observatory of Japan (NAOJ), the Kavli Institute for the Physics and Mathematics of the Universe (Kavli IPMU), the University of Tokyo, the High Energy Accelerator
Research Organization (KEK), the Academia Sinica Institute for Astronomy and Astrophysics in Taiwan (ASIAA), and Princeton University. 
Funding was contributed by the FIRST program from the Japanese Cabinet Office, the Ministry of Education, Culture, 
Sports, Science and Technology (MEXT), the Japan Society for the Promotion of Science (JSPS), Japan Science and Technology
Agency (JST), the Toray Science Foundation, NAOJ, Kavli IPMU, KEK, ASIAA, and Princeton University. 
This paper makes use of software developed for Vera C. Rubin Observatory. We thank the Rubin Observatory for making their code available as free software at http://pipelines.lsst.io/.

The Pan-STARRS1 Surveys (PS1) and the PS1 public science archive have been made possible through contributions by the Institute for Astronomy, the University of Hawaii, the Pan-STARRS Project Office, the Max Planck Society and its participating institutes, the Max Planck Institute for Astronomy, Heidelberg, and the Max Planck Institute for Extraterrestrial Physics, Garching, The Johns Hopkins University, Durham University, the University of Edinburgh, the Queen’s University Belfast, the Harvard-Smithsonian Center for Astrophysics, the Las Cumbres Observatory Global Telescope Network Incorporated, the National Central University of Taiwan, the Space Telescope Science Institute, the National Aeronautics and Space Administration under grant No. NNX08AR22G issued through the Planetary Science Division of the NASA Science Mission Directorate, the National Science Foundation grant No. AST-1238877, the University of Maryland, Eotvos Lorand University (ELTE), the Los Alamos National Laboratory, and the Gordon and Betty Moore Foundation.

\end{acknowledgments}

\facilities{CFHT (MegaCam), Subaru (Hyper Suprime-Cam)}

\software{\texttt{orbfit} \citep{ber00}, \texttt{TRIPPy} \citep{fra16},
  \texttt{MERCURY} \citep{cha99},
  and \texttt{NUMPY} \citep{Harris20}}.

\bibliography{narda}{}
\bibliographystyle{aasjournal}

\end{document}